\documentclass[final,3p,times,twocolumn]{elsarticle}

\usepackage[T1]{fontenc}
\usepackage[utf8]{inputenc}
\usepackage{amsmath}
\usepackage{amssymb}
\usepackage{arydshln}
\usepackage{epstopdf}
\usepackage{flushend}
\usepackage{hyperref}
\usepackage{graphicx}
\usepackage{lineno}
\usepackage{ulem}
\usepackage{textcomp}
\usepackage{setspace}

\usepackage{miller}

\newcommand{\etal}{\textit{et al}}
\newcommand{\ie}{\textit{i.e.}}
\newcommand{\eg}{\textit{e.g.}}

\newcommand{\abinitio}{\textit{ab initio}}
\newcommand{\Abinitio}{\textit{Ab initio}}

\usepackage{xcolor}

\journal{Acta Materialia}

\begin{document}
\begin{frontmatter}

\title{Hydrogen delaying the formation of Guinier-Preston zones in aluminium alloys}

\author[GPM]{Guillaume Hachet\fnref{Val}\corref{CA}}
\ead{guillaume.hachet@vallourec.com}
\author[GPM]{Xavier Sauvage\corref{CA}}
\ead{xavier.sauvage@univ-rouen.fr}
\cortext[CA]{Corresponding authors}

\address[GPM]{Groupe de Physique des Matériaux, Normandie University, UNIROUEN, INSA Rouen, CNRS, 76000 Rouen, France}
\fntext[Val]{Present address: Vallourec Research Center France, 60 Route de Leval, F-59620, Aulnoye-Aymeries, France}

\begin{abstract}
The consequences of hydrogen on the formation and growth of GP zones during natural ageing in an Al-5Cu alloy is investigated experimentally down to the atomic scale and numerically using \abinitio{} calculations.
As observed through scanning/transmission electron microscopy, the hardening kinetic is delayed due to a slower growth of GP zones during natural ageing when hydrogen is incorporated.
According to \abinitio{} calculations, the delayed hardening results from hydrogen trapped in vacancies that reduce significantly the diffusion coefficient of copper and the self diffusion of aluminium.
This reduction of the diffusion coefficient is either due to hydrogen being in the path of the atom that exchange with the vacancy (and increasing the energy barrier) or due to hydrogen being dissociated from the vacancy (and leading to a less stable state).
\end{abstract}


\end{frontmatter}

\section{Introduction}
\label{S1}

Aluminium alloys are widely used in industry for their various advantages: they are light weighted metallic alloys, have a high corrosion resistance and good mechanical properties \cite{Mondolfo1979,Develay1992}. 
Since the early work of Guinier \cite{Guinier1938} and Preston \cite{Preston1938}, it has been demonstrated that the increase of hardness in age hardened Al-Cu alloys is due to the formation of nanoscaled zones that have been first detected by small-angle X-ray scattering (SAXS) known as the Guinier-Preston (GP) zones \cite{Sigli2018}.
When such alloy is quenched, then aged at room temperature (\ie{}: naturally aged), the formation of these GP zones is activated by the excess vacancies after quenching, which are slowly annihilated near residual dislocations and grain boundaries.
When an Al-Cu alloy is heat treated, the precipitation sequence is \cite{Starink2004,Son2005,Rodriguez2018}:
Super Saturated Solid Solution $\rightarrow$ Cu clustering $\rightarrow$ GP zones $\rightarrow$ $\theta''$ $\rightarrow$ $\theta'$ $\rightarrow$ $\theta$.
The formation of GP zones is obtained quickly at ambient temperature, thus the copper clustering sequence is usually neglected to describe the precipitation sequence of this alloy.

GP zones in Al-Cu alloys are circular nanoscaled disks parallel to \{100\} planes, isolated in one layer \cite{Silcock1954,Bourgeois2011}.
The precipitation in Al-Cu alloys has been extensively studied in the literature through hardness measurements, transmission electron microscopy (TEM), differential scanning calorimetry (DSC) and strength models have been developed to correlate the increase of hardening to the precipitation state \cite{Myhr2001,Rodriguez2018}. 
Atomistic calculations have been used to predict the evolution during ageing of the precipitation \cite{Hutchinson2010,Sigli2018,Deschamps2021} in addition to determine the formation energy of particles, their geometries and interaction with dislocations \cite{Zander2008,Singh2010,Staab2018,Miyoshi2019}.
These parameters can be then implemented in classical nucleation and growth theories or/and clusters dynamics \cite{Clouet2009c,Stegmuller2019}.
Publications about naturally aged alloys and the role of excess vacancies are numerous, but less information is reported on GP zone nucleation and growth during the first step of natural ageing. 

Besides, aluminium alloys are among materials bearing the highest potential to contain, and to facilitate the transport of hydrogen fuel due to their various advantages and they are already used in fuel-cell-based, electric vehicles \cite{Horikawa2019}. 
More generally, since hydrogen is becoming a key component of the energy transition worldwide, it can either be used as an energy carrier or directly as a fuel in vehicles, including automobiles and planes.
However, due to its small size and high mobility, hydrogen influences the mechanical properties of metals and alloys leading to premature failures of engineering structures.
This phenomenon is called hydrogen embrittlement (HE) and several models have been proposed in the literature to describe the underlying physical mechanisms (see for details \cite{Lynch2012,Kirchheim2014,Robertson2015}).
For all models, HE involves the energy reduction of one process in the presence of hydrogen to activate a mechanism (\eg{}: grain boundary segregation \cite{Hondros1996,Oudriss2012}, Cottrell atmosphere of dislocation \cite{Gu2018,Hachet2020b}, shielding effect promoting slip band localisation \cite{Beachem1972,Birnbaum1994,Delafosse2001}, enhancement of vacancy formation \cite{McLellan1997,Harada2005,Fukai2006}, and so on...). 
These models can describe accurately HE of pure metals, but can fail describing HE in more complex materials (\eg{} in alloys) where mechanical properties are dependent on the distribution of strengthening precipitates and their interactions with defects, in particular dislocations.

Aluminium alloys are also not immune to hydrogen ingress and when hydrogen is introduced as a solute, it easily diffuses and segregates to crystalline defects \cite{Scully2012,Zhao2022}. 
\Abinitio{} calculations have shown that hydrogen atoms strongly interact with vacancies in aluminium \cite{Wolverton2004,Nazarov2014,Connetable2018}.
Previous studies focusing on the hydrogen/vacancy interactions in metals have shown that hydrogen decreases the formation energy of vacancy clusters containing hydrogen \cite{Fukai2006,Nazarov2014} and the vacancy migration energy \cite{Du2020}.
Hydrogen may also delay the clustering of solutes and the coarsening of precipitates in some Al alloys \cite{Zhao2022,Hachet2022b}. 
Therefore, it is critical to understand how these interactions impact the kinetic and thermodynamic of precipitates in aluminium alloys (which may evolve even at room temperature) to predict microstructural evolutions and eventually to reduce the damaging effect of hydrogen.
In this study, the influence of hydrogen on the early stage of GP zone formation during natural ageing in an Al-5Cu alloy is investigated. 
The following section is focused on experimental data that highlight the influence of hydrogen on GP zone nucleation and growth during the first step of natural ageing.
Then, \abinitio{} calculations are presented to first demonstrate the impact of copper on the interaction between hydrogen and vacancy.
The second part of the \abinitio{} calculations is focused on the effect of hydrogen on the diffusion of vacancy and copper in FCC aluminium. 

\section{Experimental evidence of the hydrogen influence on the GP zone formation and hardening kinetics}
\label{S2}

\subsection{Experimental details}
\label{S21}

The investigated material is provided by Goodfellow$^{\circledR}$ with the following composition (wt.\%): 5.3\%Cu-0.7\%Fe-0.4\%Si-0.3\%Pb, Al balance (standard AA2011, called Al-5Cu further). 
Disc shaped samples (with a diameter of 6\,mm and a thickness of 1\,mm) are solutionised at 810\,K during 1\,h, water quenched, then naturally aged at room temperature either in air or 5\,h in a 0.1\,M NaOH solution.
Before the introduction in the solution, samples are quickly (few minutes) mechanically grinded using SiC foil paper with a particle size of 8\,µm to remove the oxide layer grown during the solution heat treatment. 
Aqueous solution containing NaOH is aggressive towards aluminium and its oxide, it prevents the formation of a passive layer and leads to H incorporation in the alloy \cite{Birnbaum1997,Scully2012}.
After 5\,h in NaOH, the samples are further aged in air at room temperature and the hardness evolution is compared to the alloy directly aged in air.
The increasing hardness resulting from the GP zone nucleation and growth is firstly measured by micro-hardness measurements, using a Future tech FM7 device at room temperature. 
The micro-hardness values presented in this study are the average of at least 6 indents obtained with a micro Vickers diamond indenter using a load of 500\,g and a dwell time of 10\,s.
High angle annular dark field scanning TEM (HAADF-STEM) images are recorded with collection angles ranging from 50 to 180\,mrad using a JEOL ARM 200 microscope, operated at 200\,kV.
Thin foil specimens are prepared with a twin-jet electro-polisher (TENUPOL 5 from Struers$^{\circledR}$) using a mixture of $30\%\mathrm{HNO_3}-70\%\mathrm{CH_3OH}$ (\%vol) at 243\,K. 
Final thinning is carried out by low-energy ion milling conducted with a GATAN$^{\circledR}$ Precision Ion Polishing System.

\subsection{Hardness kinetic variations of the naturally aged Al-5Cu alloy due to hydrogen}
\label{S22}

\begin{figure}[bth!]
\centering
\includegraphics[width=0.99\linewidth]{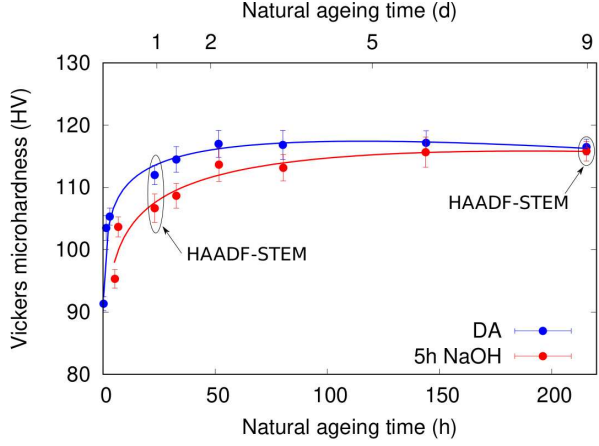}
\caption{Vickers micro-hardness evolution as a function of the natural ageing time when the alloy is diractly aged (DA) in air and after 5h in NaOH.}
\label{FigHVAgeTime_AlCu}
\end{figure}

After water quenching, the hardness of the alloy is 91$\pm$1\,HV, and it increases progressively to reach a maximum  of 117$\pm$2\,HV after $\sim50\,h$ (see fig. \ref{FigHVAgeTime_AlCu}).
When the alloy is aged 5\,h in NaOH solution to introduce hydrogen, the microhardness is significantly lower than the alloy directly aged in air: after 5\,h, the hardness of the alloy naturally aged in air is 105$\pm$2\,HV while it is only 95$\pm$2\,HV when NaOH treatment is carried out. 
However, after several additional hours (50 to 200 hours) at room temperature in air, the micro-hardness further increases and catches up the hardness of the material without hydrogen.
This suggest that hydrogen atoms quickly desorb from the alloy and do not significantly affect the final microstructure (fig. \ref{FigHVAgeTime_AlCu}).
To confirm these measurements, HAADF-STEM observations are carried out further. 

\begin{figure*}[bth!]
\centering
\includegraphics[width=\linewidth]{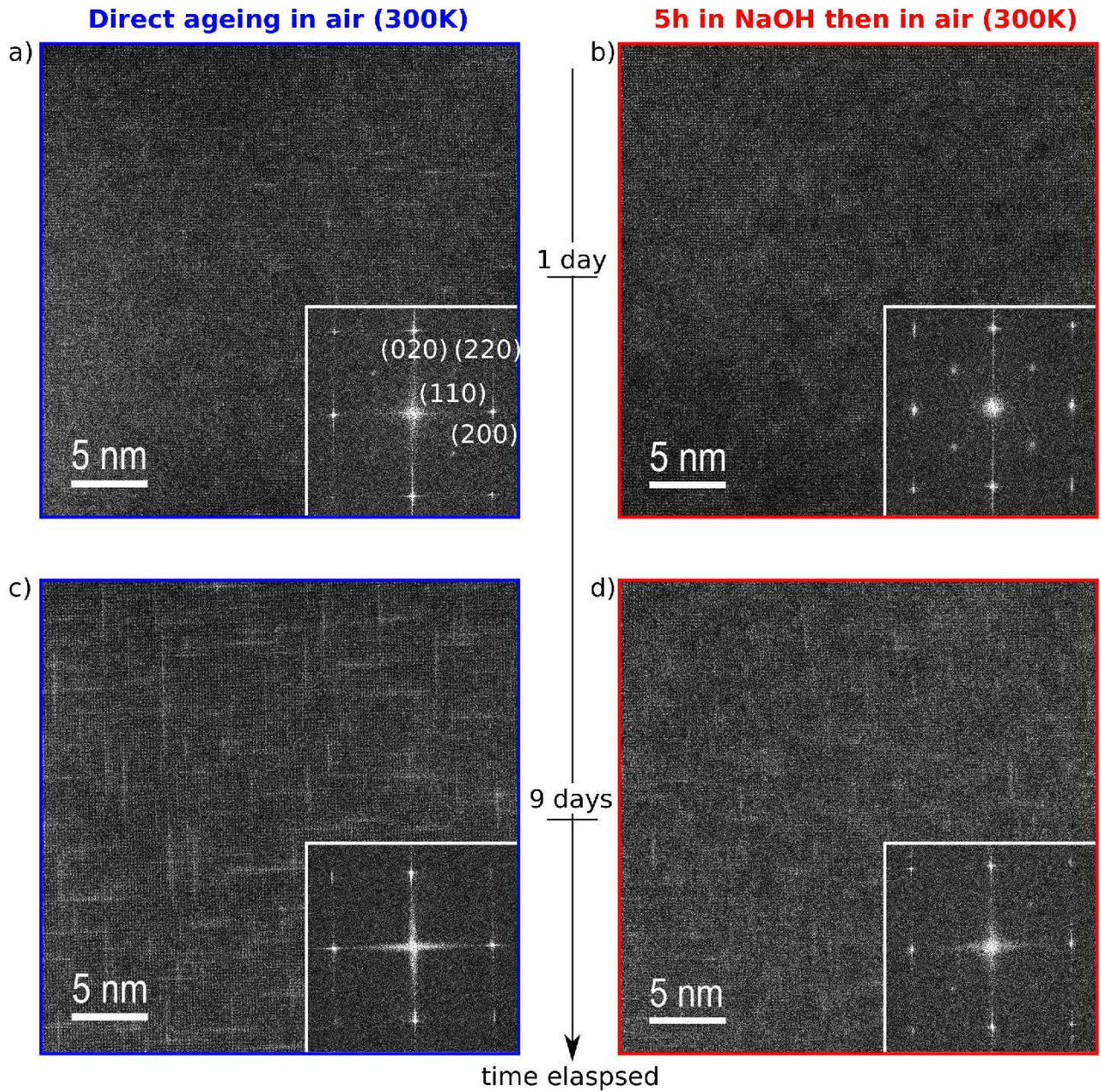}
\caption{HAADF-STEM micrographs of the precipitation evolution of the Al-5Cu alloy directly aged in air (a and c) and 5\,h in NaOH followed by an ageing in air (b and d).
The total natural ageing time is 1 day (a and b) or 9 days (c and d). 
All samples are oriented in the (001) zone axis and their corresponding FFTs are set in the bottom right corner.}
\label{FigSTEMNA}
\end{figure*}

Naturally aged materials are observed by HAADF-STEM in (001) zone axis to clearly exhibit GP zones. They are observed after being naturally aged 1 and 9 days.
When the alloy is directly aged 1 day in air, small GP zones appear, as shown in fig. \ref{FigSTEMNA}.a and they become significantly larger after 9 days (fig. \ref{FigSTEMNA}.c).
When it is aged 5\,h in NaOH, GP zones are not visible after 1 day (fig \ref{FigSTEMNA}.b), but become visible after 9 days (fig \ref{FigSTEMNA}.d), with an average diameter similar to those observed in the alloy directly aged 1 day in air (fig \ref{FigSTEMNA}.a).
These observations are consistent with the delayed hardening (fig. \ref{FigHVAgeTime_AlCu}).
However, the hardnesses of the alloy aged in both conditions are relatively similar after 9 days (fig. \ref{FigHVAgeTime_AlCu}), suggesting that hydrogen does not affect the final microstructure, which obviously is not the case when fig \ref{FigSTEMNA}.c and \ref{FigSTEMNA}.d are compared. 

\begin{figure}[bth!]
\centering
\includegraphics[width=\linewidth]{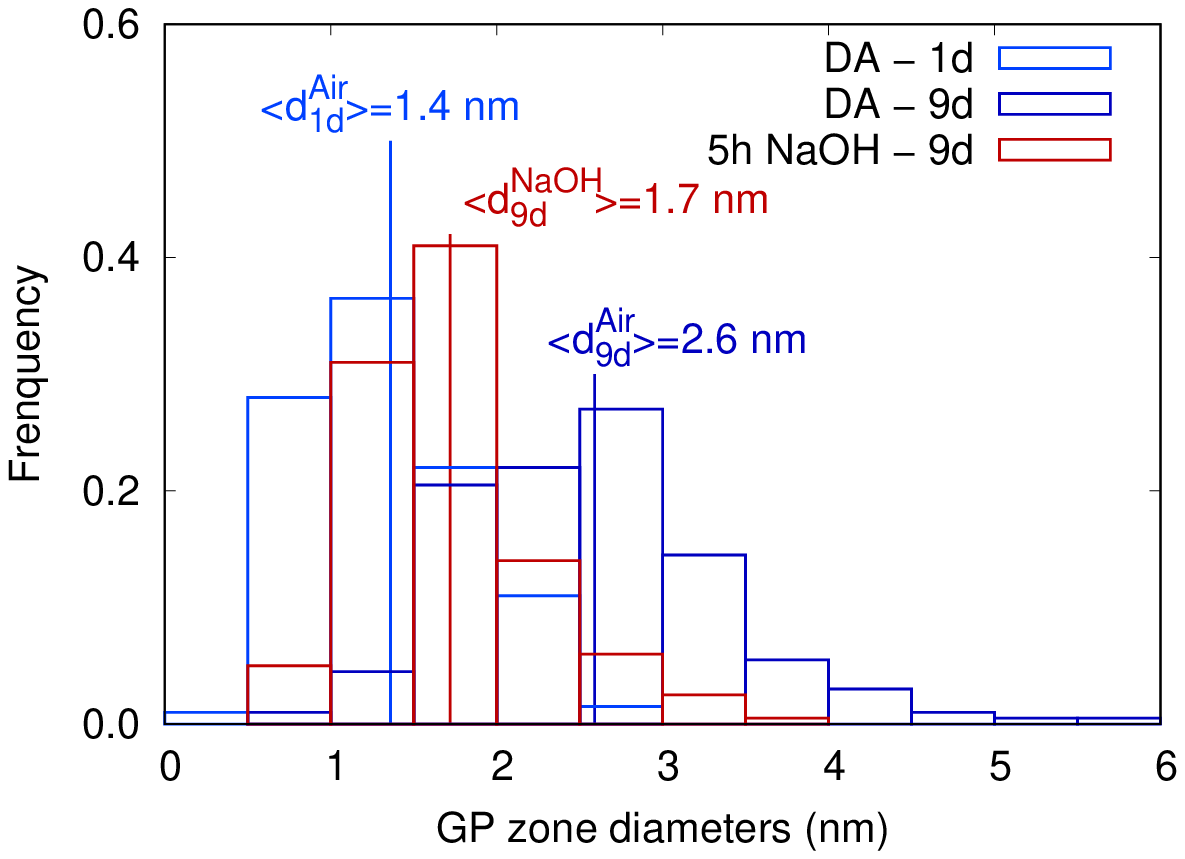}
\caption{Distribution of GP zone diameters measured on HRSTEM images in alloy directly aged in air during 1 day and 9 days (DA) and 5\,h in contact of hydrogen, then aged in air (5\,h NaOH).
The measurements were carried out on of 200\,GP zones for each distribution.}
\label{FigdGPZ}
\end{figure}

GP zone diameters $d$ are directly measured on STEM-HAADF images and their distributions are plotted in fig. \ref{FigdGPZ}.
There is no distribution for the alloy stored in 5\,h in NaOH, then aged 1 day in air because GP zones could not be observed (fig. \ref{FigSTEMNA}.b).
Longer natural ageing gives larger GP zones with a mean diameter varying form 1.4\,nm$\pm$0.5\,nm after 1 day to 2.6\,nm$\pm$0.8\,nm after 9 days.
Besides, the diameter distribution becomes significantly broader during natural ageing.
When the alloy is stored 5\,h in NaoH prior to 9 days in air, the mean diameter is 1.7\,nm$\pm$0.5\,nm with a narrow distribution similar to that of the alloy aged during 1 day directly in air.

Due to the limited foil thickness, some GP zones are partially imaged and the true diameter $d_t$ of GP zones can be determined from the apparent diameter $d$ using \cite{Nie2008,Rodriguez2018}: 
\begin{equation}
	d = \left(\frac{\pi d_t/4 + \delta}{\delta + d_t} \right)d_t,
\label{eqdapp}
\end{equation}

with $\delta$ the thin foil thickness.
In this work, thin foil thickness $\delta$ was measured by electron energy loss spectroscopy.
Assuming an electron mean free path in aluminium of 120\,nm \cite{Bardal2000}, the thickness of thin foils was systematically between 40\,nm and 100\,nm.

The density of GP zones may be also estimated, following \cite{Nie2008,Rodriguez2018}: 
\begin{equation}
	n_P = \frac{N_x + N_y + N_z}{S (d_t + \delta)},
\label{eqnP}
\end{equation}

with $N_x$, $N_y$ $N_z$, the number of GP zones counted in the x, y and z axes, parallel to the $[100]$, $[010]$ and $[001]$ directions, respectively and $S$ the surface of observation ($S=5.10^{-3} {\rm \mu m}$). 
Since only GP zones perpendicular to (001) Al planes are visible on HRSTEM images (fig. \ref{FigSTEMNA}), the number of precipitates in the z-direction ($N_z$) is obtained from $N_x$ and $N_y$ using:
\begin{equation}
	N_z = 0.5 \times (N_x + N_y)\left(\frac{d_t+\delta}{\sqrt{S}}\right).
\label{eqNz}
\end{equation}

Assuming that the thickness of all GP zones is half the aluminium lattice parameter ($a_{Al} = 0.404\,{\rm nm}$), their volume fraction writes as:
\begin{equation}
	f_V = n_P \frac{\pi d_t^2 a_{Al}}{8}.
\label{eqfV}
\end{equation}

\begin{table*}[h!]
	\centering
	\caption{Quantitative measurements of the GP zone and cluster densities $n_P$, apparent diameter $d$ and volume fraction $f_V$ in the Al-5Cu alloy directly aged in air or after 5h in NaOH.}
	\label{tblSTEMNA}
\begin{tabular}{cccccc}
\hline
\ &\  & \multicolumn{2}{c}{1 day} & \multicolumn{2}{c}{9 days}  \\
\ & \ & Air & NaOH & Air & NaOH \\
\hdashline
GP & $<d>$ (nm) & 1.4$\pm$0.5 & \ & 2.6$\pm$0.8 & 1.7$\pm$0.5 \\
zones & $n_P$ (10$^{-3}$ nm$^{-3}$) & 2.4$\pm$0.1 & \ & 3.6$\pm$0.1 & 1.6$\pm$0.1 \\
\ & $f_V$ (\%) & 0.1 & \ & 0.4 & 0.1 \\
\hdashline
Cu & $<d>$ (nm) & 1.4$\pm$0.5 & 1.3$\pm$0.4 & \ & 1.1$\pm$0.4 \\
clusters & $n_P$ (10$^{-3}$ nm$^{-3}$) & 0.8$\pm$0.1 & 1.8$\pm$0.1 & \ & 1.1$\pm$0.1 \\
\ & $f_V$ (\%) & 0.06 & 0.12 & \ & 0.04 \\
\hline
\end{tabular}
\end{table*}

\begin{figure}[bth!]
\centering
\includegraphics[width=\linewidth]{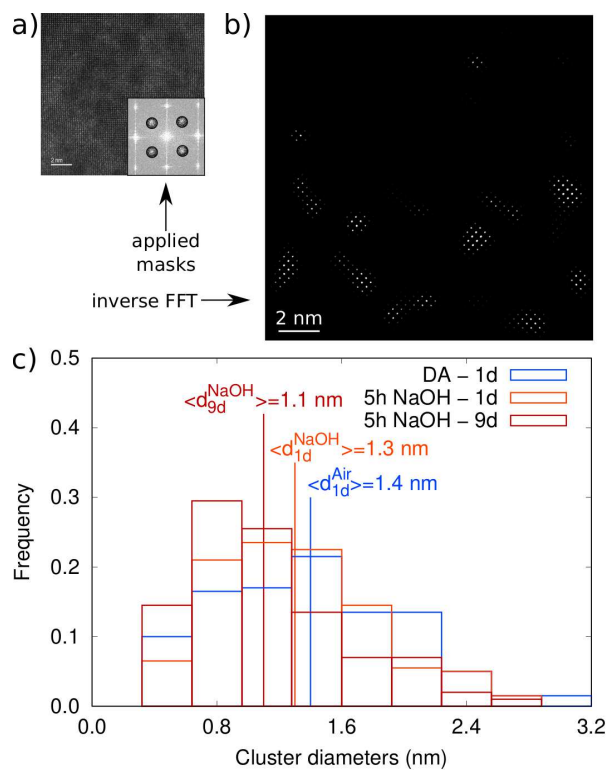}
\caption{Determination and distribution of the copper clusters' size. 
a. HAADF-STEM image of Al-5Cu aged 5\,h in NaOH and 1 day in air with the masks applied on the corresponding FFT.
b. Inverse FFT image showing regions with strong \{110\} signal on FFT. 
c. Distribution of the cluster diameters measured on filtered images corresponding to \ref{FigSTEMNA}.a, \ref{FigSTEMNA}.b and \ref{FigSTEMNA}.c.
The measurement were carried out 200 clusters for each distribution.}
\label{FigCuClDistri}
\end{figure}

Fast Fourier transforms (FFTs) of each HAADF images are also displayed in fig \ref{FigSTEMNA}.  
Spots in FFTs at the \{110\} positions are clearly exhibited for the alloy directly aged during 1 day, but they disappear after 9 days when GP zones are well developed.
These spots also appear for the alloy aged 5\,h in NaOH, followed by 1 day in air with a stronger intensity as compared to the directly aged alloy.
These spots do not completely disappear after 9 days and thus seem to be linked to features appearing before GP zones.
\{110\} reflections are forbidden reflections on diffraction pattern and they appear only on FFTs of HRSTEM images.
Similar reflections on FFT patterns have been reported in naturally aged Al-Zn-Mg alloys.
Authors have attributed these features to either $\rm{Al_3Zn}$ dispersoids or artefacts caused by the TEM sample preparations \cite{Lervik2021}.
Since we do not have such dispersoids and since similar preparation conditions were applied for all thin foils, these spots can only be the result of real features revealed by images.
Since they disappear when GP zones are well developed, they might results from Cu clusters that form prior to GP zones \cite{Starink2004,Son2005,Rodriguez2018}. 
To study these clusters, a mask is applied on the FFTs of each image having strong signals at the \{110\} positions (fig. \ref{FigCuClDistri}.a), isolating these spots. 
Then, a reconstructed image showing areas responsible of the strong \{110\} signals is obtained using an inverse FFT function (fig. \ref{FigCuClDistri}.b)
Assuming that these areas correspond to Cu clusters, their sizes are measured and the obtained distributions are displayed on fig \ref{FigCuClDistri}.c.
Contrary to GP zones, these nanosized particles are not detectable without this image filtering and disappear when the alloy is directly aged in air during 9 days (no signal is observed at the \{110\} positions of the FFTs of fig \ref{FigSTEMNA}.c).
The precipitate density is directly obtained by counting the observed clusters.
The cluster size is estimated from the average of the longest and shortest distances measured on images.
The obtained distributions are similar for all states but the average diameter is significantly smaller for samples stored in 5\,h NaOH

The determined GP zone and Cu cluster densities $n_P$, apparent diameters $d$ and volume fraction $f_V$ are summarised in table \ref{tblSTEMNA}.

These experiments clearly indicate a significant change of microstructure due to hydrogen incorporation: there is a delay of GP zone formation, observable using microhardness and HAADF-STEM.
Complementary information is detailed in \ref{AppDSC} where DSC measurements have been performed on this alloy differently aged.
Hydrogen strongly interacts with excess vacancies \cite{Fukai2006,Connetable2018,Du2020,Hachet2022b}, and thus seems to delay the formation of GP zones probably by reducing the diffusion of copper or/and by increasing the energy barrier of GP zones nucleation.
\Abinitio{} calculations are then carried out to get a deeper understanding of fundamental mechanisms leading to the delayed formation of GP zones when hydrogen is introduced on the alloy.

\section{Influence of copper atoms on the hydrogen-vacancy interactions in aluminium}
\label{S3}

\Abinitio{} calculations are performed to quantify the variation of copper diffusion when the atom is close to a hydrogen vacancy complex.
The first step is to estimate the interaction enthalpy between vacancy and hydrogen $H^{inter}_{Vac-H}$ in aluminium and in a diluted Al-Cu alloy. 
The aim is to determine if this complex is more stable in the vicinity of copper atom before determining the influence of H on the diffusion variation of copper.

\subsection{Computational details}
\label{S31}

The following \abinitio{} calculations used density functional theory (DFT) in the Quantum Espresso code \cite{Hohenberg1964,Kohn1965,Giannozzi2009}.
Pseudopotentials built with the projected augmented wave method \cite{Blochl1994,Kresse1996} are used with a kinetic energy cutoff of 600\,eV. 
The exchange-correlation is described with the generalised gradient approximation with the Perdew-Burke-Ernzerhof functional \cite{Perdew1996}.
All calculations are performed at constant pressure with a 0.2\,eV Methfessel-Paxton broadening \cite{Methfessel1989}.
All simulations cells are 4$\times$4$\times$4 repetition of the primitive cell and the Brillouin zone was sampled using a 8$\times$8$\times$8 Monkhorst and Pack \cite{Monkhorst1976}.  
Atomic positions are relaxed until all ionic forces are inferior to 10\,meV/\AA{} for static and for climbing nudged elastic band (C-NEB) calculations.
With these parameters and using the method described in previous study \cite{Hachet2018}, the obtained lattice parameters and elastic constants are deduced and given in table \ref{tblaCijAlCu}.
The experimental values are also given and showing that these parameters describe accurately both metals at atom scales \cite{Kittel2004}.

\begin{table}[h!]
	\centering
	\caption{Lattice parameters and elastic constants of pure Al and Cu.}
	\label{tblaCijAlCu}
\begin{tabular}{ccccc}
\hline
\  & $a$ (nm) & $C_{11}$ (GPa) & $C_{12}$ (GPa) & $C_{44}$ (GPa) \\
\hline
Al  & 0.40394 & 116 & 61 & 33 \\
\cite{Kittel2004} & 0.405 \ & 114 & 62 & 28 \\
\hline
Cu  & 0.36304 & 177 & 124 & 80 \\
\cite{Kittel2004} & 0.361 & 176 & 125 & 82 \\
\hline
\end{tabular}
\end{table}

\subsection{Interaction between hydrogen and vacancy in the vicinity of copper}
\label{S32}

\begin{figure}[bth!]
\centering
\includegraphics[width=0.99\linewidth]{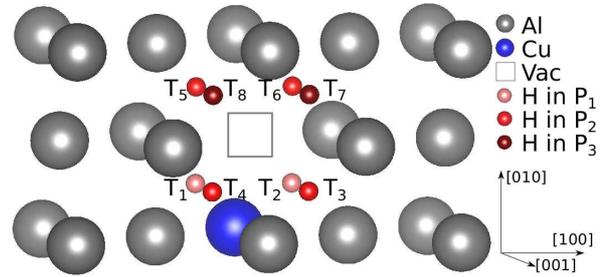}
\caption{Illustration of the hydrogen possible positions near vacancy in a diluted Al-Cu. 
The vacancy in the Al matrix (grey) containing copper (blue) is represented with the grey square.
The different possible positions for hydrogen atoms are represented with various shades of red.}
\label{FigdCuH}
\end{figure}

The first calculations aim at determining the interaction enthalpy of hydrogen vacancy complex $H_{Vac-H}^{inter}$ with and without Cu atom in its vicinity.
This enthalpy determined at zero pressure ($p=0$) is defined by:
\begin{equation}
	H^{inter}_{Vac-H} = H^{sp}_{Vac-H}(p) - H^{sp}_{Vac}(p)  - H^{sp}_{H}(p) + H^{sp}_{Bulk}(p),
\label{eqHinter}
\end{equation}

with  $H^{sp}_{Vac-H}$ and  $H^{sp}_{Vac}$, the enthalpies of the supercell containing the hydrogen-vacancy complex and the hydrogen free vacancy, respectively.
The enthalpies $H^{sp}_{H}$ and $H^{sp}_{Bulk}$ are from the vacancy free systems with and without hydrogen in solid solution, respectively.
Since the tetrahedral interstitial sites are the most stable sites for hydrogen near vacancy in pure Al \cite{Wolverton2004,Nazarov2014,Connetable2018}, we introduced hydrogen in these sites for pure Al and in a diluted Al-Cu system.
For the latter, the different interstitial sites ($T_i$ sites) are not equivalent when the copper atom is in the vicinity of the vacancy as shown in fig. \ref{FigdCuH}. 
However, due to the crystal symmetry, the distance between Cu and H when H is in $T_1$ is identical to the distance between Cu and H when H is in $T_2$. 
These equivalent positions are called $P_1$ (with $d_{Cu-H}^{P_1}$=0.04\,nm).
The distance between Cu and H is also identical when H is in $T_3$, $T_4$, $T_5$ and $T_6$, these equivalent positions are called $P_2$ ($d_{Cu-H}^{P_2}$=0.08\,nm). 
Finally, the position $P_3$ stands for H in sites $T_7$ and $T_8$ ($d_{Cu-H}^{P_3}$=0.11\,nm).
Therefore, $H^{inter}_{Vac-H}$ has to be calculated for only 3 positions in a diluted Al-Cu system.  

\begin{figure}[bth!]
\centering
\includegraphics[width=0.99\linewidth]{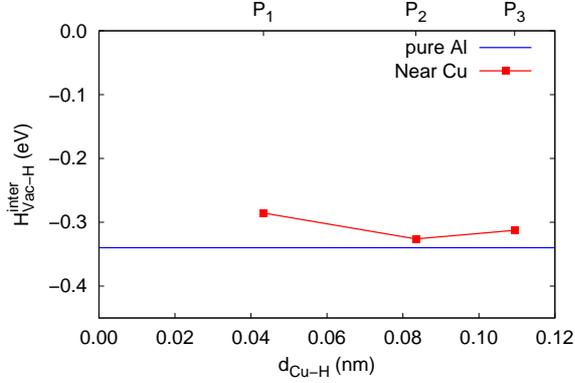}
\caption{Interaction enthalpy between vacancy and hydrogen in pure Al and in diluted Al-Cu system. 
For diluted Al-Cu, energies are displayed as a function of the distance between Cu and H atoms.}
\label{FigEinterVacH}
\end{figure}

The interaction enthalpies between hydrogen and vacancy in pure aluminium and in a diluted Al-Cu are displayed in fig. \ref{FigEinterVacH}. 
The interaction enthalpy is always attractive when hydrogen is close to the vacancy, both in pure Al and in diluted Al-Cu.
In pure aluminium, $H^{inter}_{Vac-H} = -0.34$\,eV, similar to the interaction energy reported in previous work using \abinitio{} calculations ($E^{inter}_{Vac-H} = -0.335$\,eV \cite{Wolverton2004} and $E^{inter}_{Vac-H} = -0.33$\,eV \cite{Connetable2018}) but lower than experimental data ($-0.45\pm0.07$\,eV \cite{Linderoth1988}).
In diluted Al-Cu, $H^{inter}_{Vac-H}$ is also negative, meaning that vacancies attract hydrogen, but the attraction is stronger when the vacancy is not bound to a Cu atom.
Small variations are also observed depending on the position of H in the lattice: if H is inserted in $P_1$ (\ie{}: sites $T_1$ and $T_2$ in fig. \ref{FigdCuH}), the attraction is the weakest with $H^{inter}_{Vac-H}(P_1) = -0.29$\,eV.
If H is inserted in $P_3$ (\ie{}: sites $T_7$ and $T_8$ in fig. \ref{FigdCuH}), $H^{inter}_{Vac-H}(P_3) = -0.31$\,eV and the attraction is the strongest when H is in $P_2$ position  (\ie{}: sites from $T_3$ to $T_6$ in fig. \ref{FigdCuH}), with $H^{inter}_{Vac-H}(P_2) = -0.33$\,eV. 
These interaction enthalpy variations show that copper atoms have an influence on the interaction between hydrogen atoms and vacancies in FCC aluminium matrix, but hydrogen still segregates near vacancies even with a copper atom in its vicinity and the most stable sites for H corresponds to $P_2$ and $P_3$ positions.

\section{Atomic scale modelling of hydrogen consequences on the diffusion of copper in aluminium}
\label{S4}

Further, the diffusion coefficient of copper in aluminium is evaluated using the five jump frequency model from LeClaire \cite{LeClaire1978}, which has been used in previous studies to describe the diffusion of vacancies in aluminium matrix \cite{Yang2021} or substitutionnal solute (including copper) in different matrix, including aluminium \cite{Mantina2009a,Mantina2009b,Naghavi2017}.
The first step to determine $D_{Cu}$ is to calculate the self diffusion coefficient for Al, $D_{Al}$ and the impact of hydrogen on it. 
Therefore, the influence of H on the self-diffusion coefficient of $D_{Al}$ is firstly investigated then the consequence of H of $D_{Cu}$ is studied.

\subsection{Hydrogen impact on self diffusion in pure aluminium}
\label{S41}

Since the interaction between vacancy and hydrogen is attractive in diluted Al-Cu, the self-diffusion coefficient for Al ($D_{Al}$) and the diffusion of copper ($D_{Cu}$) near hydrogen vacancy complex is determined, further. 
The self diffusion coefficient $D_{Al}$ can be defined as \cite{LeClaire1978}:
\begin{equation}
	D_{Al}(T) = f_0\omega_0(T)a(T)^2C_{0}^{j}(T),
\label{eqDAl}
\end{equation}

with $f_0$ a correlation factor, which is constant and equal to 0.7815 for FCC crystals, $a(T)$ the lattice parameter of aluminium determined at finite temperature following the method described in \cite{Metsue2016,Hachet2018} by considering the atomic vibrations of aluminium atoms (see \ref{AppLatParamDet} for more details on its determination at finite temperature).
The concentration $C_{0}^{j}(T)$ (with $j = Vac$ or $Vac-H$) is the equilibrium concentration of vacancies or hydrogen-vacancy complexes, respectively and defined as \cite{Metsue2014a,Naghavi2017}:
\begin{equation}
	C^{Vac}_0(T) = \exp\left(-\frac{H_{Vac}^f - TS_{Vac}^f}{k_BT} \right),
\label{eqCVac}
\end{equation}

where $k_B$ and $T$ are the Boltzmann constant and the temperature of the system, respectively. 
$H^f_{Vac}$ and $S^f_{Vac}$ are the vacancy formation enthalpy and entropy written in the hydrogen free system as:  
\begin{align}
	H_{Vac}^f  &= H^{sp}_{Vac} - \frac{N-1}{N} H^{sp}_{Bulk}, \label{1}\\
	S_{Vac}^f  &= S^{sp}_{Vac} - \frac{N-1}{N} S^{sp}_{Bulk}, \label{2}
\end{align}

with $H^{sp}_{Vac}$, $S^{sp}_{Vac}$, $H^{sp}_{Bulk}$ and $S^{sp}_{Bulk}$ the enthalpy and entropy of the supercell of a perfect crystal with and without a vacancy, respectively.
The obtained formation enthalpy $H_{Vac}^f$ is equal to 0.64\,eV, in agreement with the literature \cite{Mantina2008,Vo2017,Connetable2018}.
In previous work, Naghavi \etal{} have shown how the self diffusion coefficient of Cobalt is impacted by modifying $S^f_{Vac}$ \cite{Naghavi2017}. 
Hence, we have explicitly calculated $S^{sp}_{Vac}$ and $S^{sp}_{Bulk}$ with the PHON program \cite{Alfe2009}, which calculates force constant matrices and phonon frequencies in both crystals.
For hydrogen vacancy complexes, their equilibrium concentration $C^{Vac-H}_0(T)$ is:
\begin{equation}
	C^{Vac-H}_0(T) = \exp\left(-\frac{H_{Vac-H}^f - TS_{Vac-H}^f}{k_BT} \right),
\label{eqCVac-H}
\end{equation}

assuming that $S^f_{Vac-H}=S^f_{Vac}$ and with $H^f_{Vac-H}$ defined as \cite{Metsue2016,Connetable2018}:  
\begin{equation}
	H_{Vac-H}^f  = H^{sp}_{Vac-H} - \frac{N-1}{N} H^{sp}_{Bulk} - 0.5\times H_{H_2},
\label{eqHfVac_HH}
\end{equation}
with $H_{H_2}$ the enthalpy of the hydrogen molecule at zero pressure $p = 0$ MPa as a reference.
It leads to $H^f_{Vac-H}=1.05$\,eV, close to data reported in the literature (1.32\,eV \cite{Xie2016} and 1.02\,eV \cite{Connetable2018}).

The jump frequency $\omega_0(T)$ is defined as \cite{LeClaire1978}:
\begin{equation}
	\omega_{0} = \nu_0\exp\left(-\frac{H_{0}^m}{k_BT} \right),
\label{eqw0}
\end{equation}

with $\nu_0$ the attempt frequency equal to $\sqrt(3/5)k_B/h \theta_D$ ($h$ and $\theta_D$ are the Planck constant and Debye temperature of aluminium, respectively) \cite{Satta1998,Kittel2004}. 
$H_{0}^m$ is the migration enthalpy of vacancies or hydrogen vacancy complexes.
Since all calculations are performed at zero pressure, it is assumed that the migration enthalpy $H_{0}^m$ is equivalent to the migration energy $E^m_{0}$ \cite{Carling2003,Naghavi2017}, which is obtained from the transition states of the C-NEB calculations of vacancy (with or without H) diffusing from one site to another.

\begin{figure}[bth!]
\centering
\includegraphics[width=\linewidth]{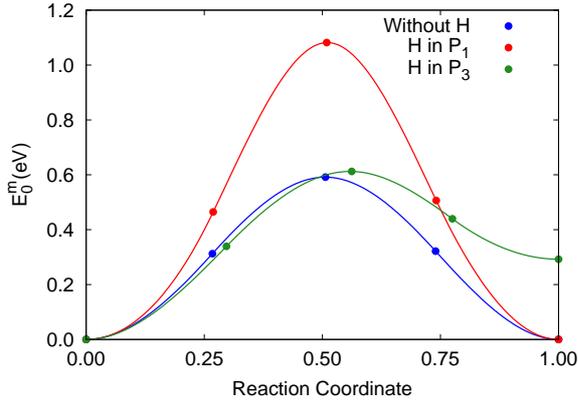}
\caption{Energy variation of Al atom migrating in pure Al near vacancy and hydrogen-vacancy complex.}
\label{Figw0}
\end{figure}

Without H, the vacancy migration enthalpy presented in fig. \ref{Figw0} is found $H_{Vac}^m=E_{Vac}^m=0.59$\,eV, in agreement with the literature \cite{Mantina2008}.
Assuming Al atoms exchange a similar way with the vacancy in pure Al and in a diluted Al-Cu, hydrogen can either be still associated to the vacancy (when H is in $P_1$) or dissociated from the vacancy (when H is in $P_3$).
The migration energy of a hydrogen-vacancy complex (with H in $P_1$ and $P_3$) is also obtained using C-NEB calculations and results are also presented in fig. \ref{Figw0}.
When H is still associated to the vacancy during the Al atom displacement, an increase of the migration energy is observed ($E_{Vac-H}^m(P_1)=1.08$\,eV) while $E_{Vac-H}^m(P_3)=0.61$\,eV, close to $E_{Vac}^m$.
However, the final state is less stable when H is in $P_3$, the energy difference corresponding to the hydrogen-vacancy binding energy.

\begin{figure}[bth!]
\centering
\includegraphics[width=\linewidth]{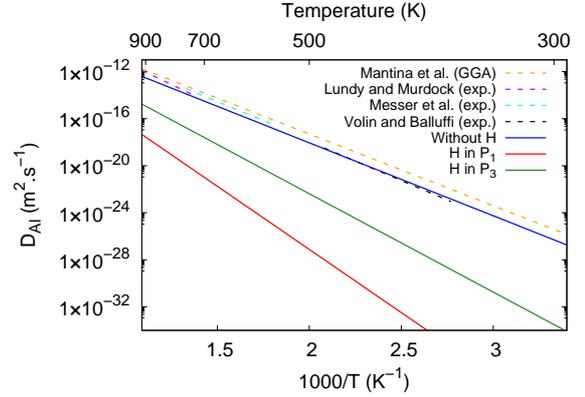}
\caption{Calculated self-diffusion coefficient for Al, including previous work of Mantina \etal{} \cite{Mantina2008}, Lundy and Murdock \cite{Lundy1962}, Messer \etal{} \cite{Messer1975}, Volin and Ballufi \cite{Volin1968}}.
\label{FigDAl}
\end{figure}

The self-diffusion coefficient calculated as a function of the temperature is displayed in fig. \ref{FigDAl}.
The coefficient $D_{Al}$ obtained from our \abinitio{} calculations is close to the values reported in the literature in a range of 360\,K to 933\,K \cite{Mantina2008,Lundy1962,Volin1968,Messer1975}.
In presence of H, $D_{Al}$ is always reduced but the diffusion coefficient is less impacted when H is in $P_3$ than when it is in $P_1$ (due to the increase of migration energy $E_{Vac-H}^m(P_1) = 1.08$\,eV).

The diffusion of impurities or vacancies usually follows an Arrhenius type temperature dependence \cite{LeClaire1978}: 
\begin{equation}
	D_{Al}(T)  = D_0 \exp{\left(- \frac{Q}{k_B T}\right)},
\label{eqDAlArr}
\end{equation}

with $D_0$ a pre-exponential factor and $Q$ the activation energy. 
From \abinitio{} calculations (fig. \ref{FigDAl}), curves are fitted using eq. (\ref{eqDAlArr}) to quantify the change of $D_0$ and $Q$ due to hydrogen incorporation.
The results are given in table \ref{tblD0QDAl}.
Without hydrogen, calculated data are similar to those reported in the literature but an important increase of the activation energy is obtained when hydrogen-vacancy complexes exist.  

\begin{table}[h!]
	\centering
	\caption{Pre-exponential factor $D_0$ and activation energy $Q$ of self-diffusion coefficient for Al with and without H.
	A comparison with data from the literature is also provided \cite{Mantina2008,Lundy1962,Volin1968,Messer1975}.
	* are for experimental data and ** are for \abinitio{} calculations}
	\label{tblD0QDAl}
\begin{tabular}{cccc}
\hline
\  & $D_0$ ($10^{-6}$ m$^2$.s$^{-1}$) & $Q$ (eV) & $T$ (K) \\
\hline
* \cite{Lundy1962} & 171 & 1.48 & 720-920 \\
* \cite{Volin1968} & 17.6 & 1.31 & 360-480 \\
* \cite{Messer1975} & 13.7 & 1.28 & 550-750 \\
** \cite{Mantina2008} & 7.75 & 1.26 & \ \\
Without H  & 2.41 & 1.23 & \ \\
With H in $P_1$ & 2.41 & 2.13 & \ \\
With H in $P_3$ & 2.41 & 1.66 & \ \\
\hline
\end{tabular}
\end{table}

Although the interaction enthalpy between a vacancy and a H atom is negative, these calculations show that in pure Al, hydrogen delays the diffusion of vacancy by increasing the vacancy migration enthalpy (when H is in $P_1$ only) and vacancy formation enthalpy (when H is in $P_1$ and $P_3$). 
Thus, hydrogen should also impact the diffusion of copper, as studied in the next section. 

\subsection{Copper diffusion in the vicinity of hydrogen-vacancy complexes in Al matrix}
\label{S42}

The influence of hydrogen on the diffusion coefficient of copper ($D_{Cu}$) is investigated in this section using the five jump frequency model developed by LeClaire \cite{LeClaire1978}.
The model includes several assumptions: (i) the solute diffusion is controlled by a vacancy mechanism, (ii) the interaction between solute and vacancy is only limited to monovacancy being in the nearest neighbour of the solute and (iii) Cu atoms do not interact with other Cu atoms (diluted system).
It has been demonstrated that such model accurately describes the diffusion of solute in aluminium \cite{Mantina2009a,Mantina2009b} or in cobalt \cite{Naghavi2017}.
This approach considers that solute atoms diffuse predominantly with a vacancy mechanisms through the five jump frequencies $\omega_i, i = 0,4$ defined by \cite{LeClaire1978}: 
\begin{equation}
	\omega_i(T) = \nu_i\exp{-\frac{H_{i}^m}{k_BT}}.
\label{eqwi}
\end{equation}

$\omega_1$ corresponds to the jump frequency for Al atom-vacancy jumps between a pair of sites that are both nearest neighbours of Cu atom.
$\omega_2$ is the jump frequency for Cu atom-vacancy exchange.
$\omega_3$ is the jump frequency of Al, which dissociates the vacancy and Cu.
$\omega_4$ is the opposite of $\omega_3$ (\ie{}: jump frequency of Al atom, which binds Cu and the vacancy). 
The fifth jump is $\omega_0$, the self-diffusion of Al (see eq. (\ref{eqw0}) in section \ref{S32}).
The attempt frequency $\nu_i$ is equal $\sqrt(3/5)k_B/h \theta_D$ (like in the previous section), with $\theta_D$ the Debye temperature of aluminium or copper depending on the atom that is exchanged with the vacancy.

\begin{figure}[bth!]
\centering
\includegraphics[width=0.67\linewidth]{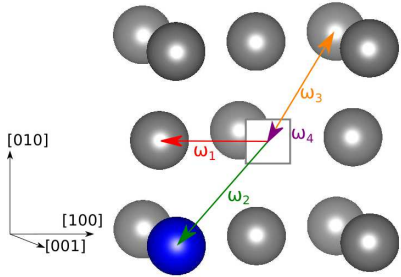}
\caption{Illustration of 4 atom jumps based on the LeClaire model (the fifth jump, $\omega_0$ correspond to the jump frequency of an Al atom without copper).}
\label{FigCuVac}
\end{figure}

NEB calculations are performed to determine the migration energies of the different jumps without H and with H in $P_1$ and $P_3$ positions. 
Like in section \ref{S41}, these positions are chosen over $P_2$ in order to have H either still associated with the vacancy before and after the jump or dissociated from it .
Besides, all calculations are performed at zero pressure, therefore it is also assumed that the migration enthalpies $H_{i}^m$ are equivalent to the migration energies $E^m_{i}$ \cite{Carling2003,Naghavi2017}.

\begin{figure}[bth!]
\centering
\includegraphics[width=\linewidth]{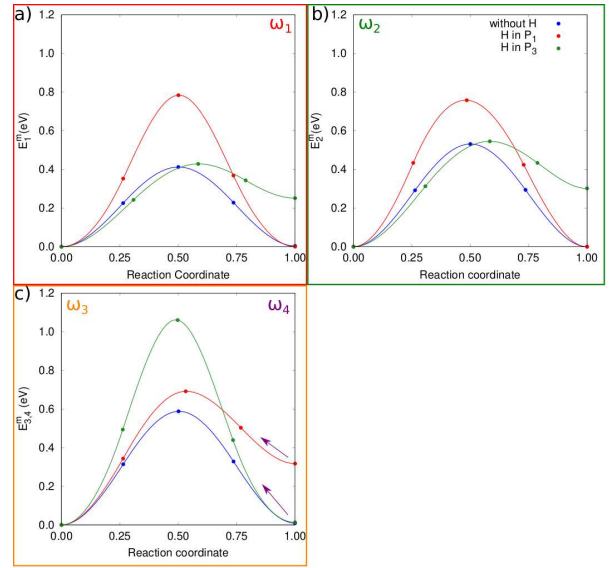}
\caption{Energy variation of atoms corresponding to the different jump frequencies for vacancy and hydrogen vacancy complexes. 
a. Energy migration of Al that does not dissociate copper with the vacancy.
b. Energy migration of Cu in Al matrix.
c. Energy migration of Al that dissociates (orange) and associates (dark-magenta) the vacancy with Cu atom.}
\label{Figwi}
\end{figure}

Results are plotted in fig. \ref{Figwi}.
When H is still associated to the vacancy, the migration energy is always larger than when the atom is exchanged with a hydrogen free vacancy.
When H is the furthest from the mobile atom, H is dissociated from the vacancy and the final state has an higher energy than its initial state corresponding to the interaction energy between H and vacancy. 
Consequently, when H is in $P_1$, $E^m_4$ is lower than $E^m_3$, while it is symmetrical when the vacancy is not linked to a hydrogen atom.

Then, the diffusion coefficient of copper $D_{Cu}$ is obtained using \cite{LeClaire1978}:
\begin{equation}
	D_{Cu}(T) = f_2\omega_2a(T)^2C_2^{j}(T),
\label{eqDCu}
\end{equation}

with $f_2$ a correlation factor, similar to $f_0$ in eq. (\ref{eqDAl}), which is written \cite{Manning1964}:
\begin{equation}
	f_2 = \frac{2\omega_1+\omega_3 F(\alpha)}{2\omega_2+2\omega_1+\omega_3 F(\alpha)},
\label{eqf2}
\end{equation}

where $\alpha=\omega_4/\omega_0$ and 
\begin{multline}
	F(\alpha) = 7 - \left(\frac{10\alpha^4+180.5\alpha^3+927\alpha^2+1341\alpha}{2\alpha^4+40.2\alpha^3+254\alpha^2+597\alpha+436}\right).
\label{eqF2}
\end{multline}

The concentration $C^{j}_2(T)$ is similar to $C^{j}_0(T)$ of eq. (\ref{eqDAl}) and is written as:
\begin{equation}
	C^{j}_{2}(T) = \exp\left(-\frac{H_{j}^f - TS_{i}^f + \Delta G_b}{k_BT} \right),
\label{eqC2}
\end{equation}

with the $\Delta G_b$, corresponding to the binding energy between copper and vacancy.
This energy is linked to the concentration of copper atoms near vacancies and can be determined using the jump frequencies ratio within the assumptions detailed in the beginning of this section \cite{LeClaire1978}:
\begin{equation}
	\frac{\omega_3}{\omega_4} = \exp{\left(-\frac{\Delta G_b}{k_B T}\right)}.
\label{eqDGb}
\end{equation}

\begin{figure}[bth!]
\centering
\includegraphics[width=\linewidth]{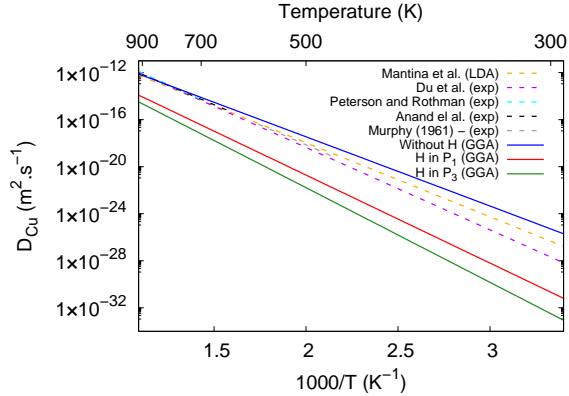}
\caption{Calculated diffusion coefficient of Cu as a function of T with and without H compared to previous work of Mantina \etal{} \cite{Mantina2009a}, Du \etal{} \cite{Du2003}, Peterson and Rothman \cite{Peterson1970}, Anand \etal{} \cite{Anand1965} and Murphy \cite{Murphy1961}}.
\label{FigDCu}
\end{figure}

Identical to section \ref{S41}, the concentration and migration energy of hydrogen-vacancy complexes are calculated for systems having hydrogen in $P_1$ and $P_3$ and the resulting coefficient diffusion of copper is plotted in fig. \ref{FigDCu}.

While a slightly higher diffusion coefficient is obtained at low temperature compared to literature data, relatively consistent values are obtained at high temperatures ($T<500\,K$). 
Moreover, the diffusion coefficient of copper is systematically lower in presence of hydrogen.
Assuming that hydrogen atoms stay in its interstitial site, the effect is more important when H is far from the copper atom (in $P_3$) than when it is close to Cu (in $P_1$). 
This result is obtained because when H is far from Cu, the energy barrier to associate the hydrogen vacancy complex ($E^m_4$) or to dissociate it with the Cu atom ($E^m_3$) is large.
Besides, when Cu atom is exchanged with the vacancy, it dissociates H-vacancy complex (which is less stable than the initial state). 
These effects have a more important impact on the diffusion coefficient than having hydrogen in $P_1$, which increases the migration energy of copper $E^m_2$. 

Assuming that the diffusion of copper also follows an Arrhenius type temperature dependence (eq. \ref{eqDAlArr}),
$D_0$ and $Q$ can also be estimated for Cu in pure Al, by fitting curves of fig. \ref{FigDCu}.
The fitted results are displayed in table \ref{tblD0QDCu}.
Without hydrogen, the calculated data are significantly lower than those reported in the literature.
An important increase of the activation energy is also obtained when hydrogen is linked to the vacancies, but the variation of the activation energy is less pronounced for $D_{Cu}$ than for $D_{Al}$.   

\begin{table}[h!]
	\centering
	\caption{Pre-exponential factor $D_0$ and activation energy $Q$ of the diffusion coefficient of Cu in FCC Al with and without H.
	A comparison with data from the literature is also provided \cite{Mantina2009a,Du2003,Peterson1970,Anand1965,Murphy1961}.
	* are for experimental data and ** are for \abinitio{} calculations.}
	\label{tblD0QDCu}
\begin{tabular}{cccc}
\hline
\  & $D_0$ ($10^{-6}$ m$^2$.s$^{-1}$) & $Q$ (eV) & $T$ (K) \\
\hline
* \cite{Murphy1961} & 29 & 1.35 & 777-908 \\
* \cite{Anand1965} & 15 & 1.31 & 623-903 \\
* \cite{Peterson1970} & 65 & 1.40 & 594-928 \\
* \cite{Du2003} & 44 & 1.39 & \ \\
** \cite{Mantina2009a} & 4.4 & 1.25 & \ \\
Without H  & 1.9 & 1.16 & \ \\
With H in $P_1$ & 1.7 & 1.48 & \ \\
With H in $P_3$ & 1.7 & 1.59 & \ \\
\hline
\end{tabular}
\end{table}

\section{Discussion} 
\label{S5}

The experiments of section \ref{S2} show a delay of the formation and growth of GP zones when hydrogen is introduced in solid solution. 
In the following section, a classical strength model is applied to correlate the microstructure evolution observed through HAADF-STEM with the hardness measurements.
Then, results from experiments are compared with the diffusion coefficients obtained in section \ref{S3} and discussed further.

\subsection{Relationship between structure evolution and hardness}
\label{S51}

Classical strengthening model is applied to establish a relationship between the hardness measurements and the HAADF-STEM observations (model developed in previous work for Al alloys \cite{Myhr2001,Bardel2014,Rodriguez2018,Bellon2020}).
It is assumed that the microhardness is linked to the yield stress $\sigma_Y$ through a Tabor factor (T = 2.8) \cite{Hutchings2009} and $\sigma_Y$ is linked to the stress needed to induce dislocations or to make them mobile (\ie{}: the critical resolved shear stress) $\tau_Y$ through a Taylor factor (M = 3.1) \cite{Bardel2014,Rodriguez2018}:
\begin{equation}
	\tau_Y = \frac{T}{M}HV.
\label{eqHVty}
\end{equation}

The stress $\tau_Y$ can be determined from microstructural features and additive contributions; which may be considered as a first approximation as \cite{Myhr2001,Bardel2014,Rodriguez2018}:
\begin{equation}
	\tau_Y = \tau_0 + \tau_d + \tau_{gb} + \tau_{ss} + \tau_P
\label{eqty}
\end{equation}

where $\tau_0$ is the friction stress, $\tau_d$ the forest hardening due to dislocations, $\tau_{gb}$ the stress contribution of the grain boundaries, $\tau_{ss}$ the stress contribution Cu of solid solution and $\tau_P$ the stress contribution of particles.
The stress contributions $\tau_0$, $\tau_d$ and $\tau_{gb}$ are assumed constant for all ageing times and they are determined from hardness measurements of the alloy naturally aged 10\,minutes, assuming that all the copper atoms are in solid solution.
The stress contribution $\tau_{ss}$ may be written \cite{Rodriguez2018,Bellon2020}:
\begin{equation}
	\tau_{ss} = HX_{Cu}^{n},
\label{eqtss}
\end{equation}

with $X_{Cu}$ the mass fraction of Cu in solid solution in the Al matrix. $H$ and $n$ two constants equal to 7.2\,MPa and 1, respectively \cite{Zhu2008,Bellon2020}.
Then, the concentration of copper in solid $X_{Cu}$ is updated for each state by knowing the volume fraction of precipitates (Cu clusters and GP zones) inducing a variation of $\tau_{ss}$.
When the alloy is naturally aged, Cu atoms agglomerate as clusters then quickly form GP zones. 
According to the HAADF-STEM images, the kinetic is slower when the alloy is in contact of hydrogen prior to ageing in air.
Both GP zones and Cu clusters affect hardness due to interactions with dislocations.
Since they are nanoscaled with a diameter $d_t$ inferior to $d_C = 10 \rm{nm}$, a critical diameter below which dislocations shear these particles (and when the particles are larger than $d_C$, they are by-passed by dislocations) \cite{Myhr2001}, the shear stress $\tau_P$ for a moving dislocation writes as \cite{Myhr2001,Rodriguez2018}:
\begin{equation}
	\tau_{P} = \frac{1}{b}\sqrt{\frac{3 f_V}{2 \pi}}\left(0.72 \mu b^2 \right) \left(\frac{2}{d_C}\right)^{1.5}\left(\frac{d_t}{2}\right)^{0.5},
\label{eqtp}
\end{equation}

with $b$ and $\mu$, the Burgers vector and shear modulus equal to 0.286 nm and 27 GPa \cite{Myhr2001}, respectively.
$d_t$ and $f_V$ are experimental data (table \ref{tblSTEMNA}) and they are used to determine $\tau_P$ for each condition.
It is important to note that eq. (\ref{eqtp}) was established for spherical particles, which is clearly not the case for GP zones.
However, previous studies have shown that this equation gives also consistent results for GP zones \cite{Zander2008,Rodriguez2018}.

\begin{figure}[bh!]
\centering
\includegraphics[width=\linewidth]{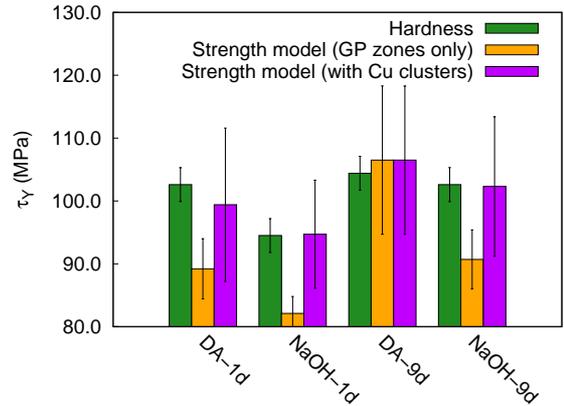}
\caption{Variation of $\tau_Y$ during natural ageing of AlCu, obtained from the hardness measurements and from the strength model considering GP zones only and including particles appearing as copper clusters.}
\label{FigTauY}
\end{figure}

Fig. \ref{FigTauY} shows the variations of $\tau_Y$ estimated from hardness measurements and eq. (\ref{eqHVty}).
These estimates are compared to values calculated from model, eqs. (\ref{eqty}-\ref{eqtp}), with only the contribution of GP zones and with the contribution of GP zones and Cu clusters.
The shear stresses obtained from the strength model exhibit the same trend than values estimated from hardness measurements: (i) $\tau_Y$ increases with the ageing time and (ii) $\tau_Y$ is always larger when the alloy is aged in air than when it is stored in NaOH prior to ageing in air.
However, the best match is obtained when both clusters and GP zones are considered.
This result confirms the delay of GP zone nucleation and growth due to hydrogen incorporation.

\subsection{Comparison between experiments and calculations} 
\label{S52}

The microhardness measurements show a delay of the hardening kinetic when the alloy is stored in 5\,h in NaOH. 
The HAADF-STEM observations confirm this delay and even after being naturally aged 9 days, microstructures are still different when the alloy is directly aged in air or stored 5\,h in NaOH.
This difference can be due to hydrogen reducing the diffusion of copper or/and an increase of the energy barrier to form GP zones, locking copper agglomerates as clusters.
Since hydrogen strongly interacts with the excess vacancies \cite{Fukai2006,Connetable2018,Du2020,Hachet2022b}, our numerical study focused on the effect of hydrogen on the diffusion of the copper in aluminium through \abinitio{} calculations.

According to fig. \ref{FigDCu}, the ratio of diffusion coefficient $D^{With H}_{Cu}/D^{Without H}_{Cu}$ is $\sim10^{-6}$ at 300\,K.
Defining the effective diffusion length of copper as $\lambda_{Cu} = \sqrt{6 D_{Cu} t}$, then $\lambda^{With H}_{Cu}/\lambda^{Without H}_{Cu} \sim 10^{-3}$.
The HAADF-STEM images of fig. \ref{FigSTEMNA} show that the structure is similar in the alloy directly aged 1\,h in air and in the alloy stored 5\,h in NaOH and aged 9 days in air. 
Assuming that this difference is only due to the impact of hydrogen on the diffusion coefficient, then $\lambda^{With H}_{Cu}(t=9\,d) = \lambda^{Without H}_{Cu}(t=1\,d)$ and $D^{With H}_{Cu}/D^{Without H}_{Cu} \sim 10^{-1}$.
This is consistent but much lower than results from \abinitio{} calculations.
However, during the first hours in NaOH, hydrogen diffuses in Al-5Cu and some time is required to create H-vacancy complexes, thus part of them annihilate before the creation of a hydrogen-vacancy complex.
Besides, during the ageing in air after the 5\,h in NaOH, hydrogen may desorb from the alloy, which probably reduces the quantity of hydrogen-vacancy complexes.
These two features may explain the difference \d{of diffusion coefficient} between experimental observations and calculations.
Nevertheless, the \abinitio{} calculations are focused only on single H-vacancy complexes while previous work has shown that depending on the hydrogen content, hydrogen-vacancy complexes with several hydrogen atoms may be more stable than single H-vacancy complexes \cite{Gunaydin2008}. 
However, these H atoms would be placed in additional tetrahedral interstitial sites of the vacancy, decreasing even more the diffusion coefficient and may lead to a non-Arrhenius behavior of the diffusion coefficient (as observed for $D_{Al}^H$ in ref. \cite{Gunaydin2008}).

\Abinitio{} calculations revealed that hydrogen and vacancy are attractive, even with copper in the vicinity of the vacancy (and H is more stable far from Cu atoms, in positions $P_2$ and $P_3$).
The previous calculations for pure Al (sec. \ref{S41}) show that hydrogen affects the mobility of vacancy and thus should also impact the diffusion of copper as confirmed in section \ref{S42}.
However, it is important to note that these \abinitio{} calculations have some limitations.
It has been established that hydrogen enhances the formation of vacancies in aluminium \cite{Fukai2006,Xie2016} even though an increase of the formation enthalpy of hydrogen-vacancy complex is obtained (which is also in agreement with the literature \cite{Xie2016,Connetable2018}). 
This is because \abinitio{} calculations focus on monovacancies bound exclusively to one hydrogen atom. 
According to previous \abinitio{} calculations, up to 10\,H atoms can be incorporated in one vacancy (12\,H atoms for \cite{Lu2005} and 13\,H for \cite{Nazarov2014}).
While increasing the number of H-trapped atoms in one vacancy may increases the formation enthalpy of the hydrogen-vacancy complex \cite{Connetable2018}, several hydrogen atoms included in one vacancy can form $H_2$ molecules.
This might reduce the formation energy of these complexes or lead to divacancies, which have a stronger attraction with H \cite{Nazarov2014}. 
In the situation where several H are trapped in vacancies or divacancies, the complex may reduce its formation energy considerably and would not impact the migration energy of the vacancy (\eg{}: when H atoms are far of the moving atom).
Theoretically, even an increase of the self-diffusion due to H can not be excluded. 
Such increase of self-diffusion has been reported for nickel where the formation energies of hydrogen-vacancy follows the superabundant vacancy model with up to 6\,H atoms inserted \cite{Metsue2016}. 
However, the increase of the migration energy of these complexes due to H incorporation is smaller \cite{Wang2015}.
As a consequence, when one hydrogen atom is incorporated in one vacancy, hydrogen delays its diffusion, but the opposite may occur when more hydrogen atoms are incorporated in one vacancy, and an increase of the vacancy mobility can even be obtained \cite{Du2020}.
Similar mechanisms cannot be excluded in the diluted Al-Cu system, but experimental data show that Cu diffusion is significantly reduced, thus based on our \abinitio{} calculations, single H-vacancy complexes are probably dominant.

\section{Conclusions}
\label{S6}

This present work studied the consequences of hydrogen on the GP zone formation and growth in an Al-5Cu alloy.
When it is naturally aged in air, excess vacancies diffuses and are annihilated in residual dislocations and grain boundaries, allowing copper to diffuse leading to the nucleation of GP zones, which that harden the alloys.
When the alloy is in contact with hydrogen during the beginning of the natural ageing, a delay of the hardening kinetics is noted with a change of the microstructure observed through HAADF-STEM.
According to the \abinitio{} calculations performed in this study, hydrogen-vacancy complexes are stable in the vicinity of Cu in a diluted Al-Cu system.
Therefore, these complexes \d{(containing one or more H atoms)} affect the diffusion of Cu in Al matrix leading to a delayed nucleation and growth of GP zones.

Modelling at atomic scale the self-diffusion for Al and the diffusion of Cu in Al in the presence of hydrogen-vacancy complexes highlights two phenomenons : (i) hydrogen in the path of the atom that exchange with the vacancy will increase the energy barrier and (ii) hydrogen far from this atom will be dissociated from the vacancy and it will be less stable after the atom jump.
As a consequence, the diffusion coefficient of copper is always smaller in presence of hydrogen close to or trapped in the vacancy.
In addition, even if hydrogen always reduces the mobility of Cu in diluted Al-Cu, it was demonstrated that the position of H in the lattice of a diluted Al-Cu alloy influences the diffusion coefficient.
Therefore, statistical study using techniques such as kinetic Monte-Carlo simulations would be performed in the near future to determine the evolution of the diffusion coefficient of copper as a function of the hydrogen concentration.
This statistical study would also help determining the number of H atoms insides a vacancy for a given temperature and H concentration, which is an unknown parameter in the present work.
Finally, these results are obtained without studying the consequences of hydrogen on the precipitate/matrix interface, which will also be the object of a future study.

\section*{Acknowledgments} 
The authors thanks Dr. D. Embury, Dr. A. Saiter-Fourcin, Dr. F. Vurpillot and G. Da Costa for fruitful discussions.
The authors acknowledge the financial support of the French Agence Nationale de la Recherche (ANR), through the program “Investissements d’Avenir” (ANR-10-LABX-09-01), LabEx EMC3, and the région NORMANDIE.
This work was partially supported by the CRNS Federation IRMA - FR 3095.
The calculations are performed using the computing ressources of CRIANN (Normandy, France) whithin the framework of the project No. 2021016.

\appendix

\section{Differential scanning calorimetry measurements of differently aged Al-5Cu}
\label{AppDSC}

DSC measurements are carried out using a Q100 TA Instruments DSC apparatus.
Samples are manufactured in 2\,mm$\times$2\,mm$\times$0.25\,mm dimensions and with a mass lower than 8\,mg.
The heat flow measurements are carried out linearly from 253\,K to 783\,K at a rate of 10\,K.min$^{-1}$.
Samples are cooled and then heated a second time from 253\,K to 783\,K to record the baseline.
Then, the heat flows presented in this work correspond to the difference between the two data set. 
GP zones are observed using TEM with a JEOL-ARM200F microscope operated at 200kV.

\begin{figure}[bth!]
\centering
\includegraphics[width=0.99\linewidth]{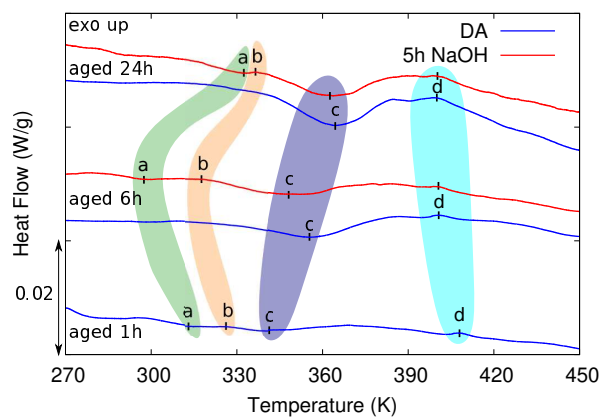}
\caption{Heat flow curves of naturally aged Al-5Cu alloy in air (blue) or 5h in NaOH (red). Peak $a$ corresponds to the dissolution of Cu clusters, peaks $b$ and $c$ the formation and dissolution of GP zones, peaks $d$ the formation of $\theta''$ \cite{Starink2004,Son2005,Rodriguez2018}.
Heat flow curves of the alloy aged 5\,h NaOH include the time spend in the aqueous solution (\ie{}: when the alloy is aged is 6\,h or 24\,h, it is aged 5\,h in NaOH then an additional 1\,h or 19\,h in air, respectively).}
\label{FigDSC_AlCu}
\end{figure}

DSC measurements are carried out to exhibit the exothermic and endothermic peaks of the precipitation sequence, which may vary due to hydrogen.
The heat flow curves of five samples of Al-5Cu alloy differently aged are displayed fig. \ref{FigDSC_AlCu}.
Since, the time needed to prepare samples after the heat treatment is about 1\,h, the heat flow curve of the material aged 1\,h in air may be considered as the as-quenched state.
In the DA alloy, the endothermic peak corresponding to the dissolution of the copper clustering and the exothermic peak corresponding to the formation of GP zones are observed (peaks $a$ and $b$ in fig. \ref{FigDSC_AlCu}) but disappear after when the material is aged 6\,h.
It is obtained due to the excess vacancies lead to the formation of GP zones at ambient temperature.
When the alloy is aged 5\,h in NaOH, these peaks seem to be still observed even up to 24\,h.
This result is in agreement the one obtained using micro-hardness measurements and HAADF-STEM observations: hydrogen affects the microstructural evolution of naturally aged Al-5Cu alloys by reducing nucleation and growth kinetics of GP zones.

\section{Lattice parameter value at finite temperature}
\label{AppLatParamDet}

\begin{figure}[bth!]
\centering
\includegraphics[width=\linewidth]{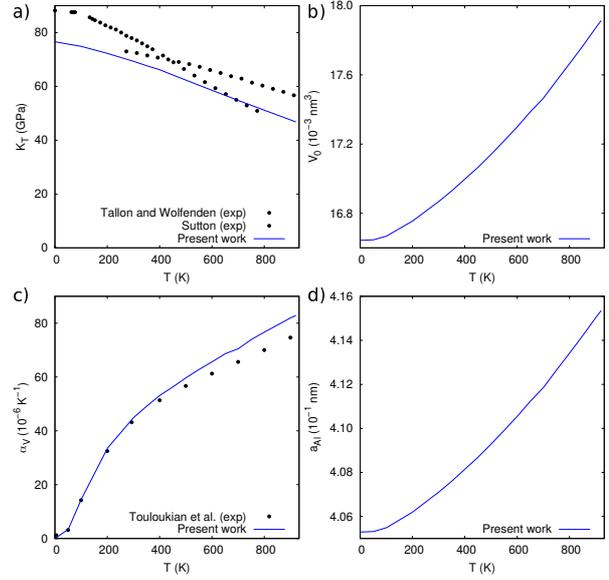}
\caption{Determination of the lattice parameter at finite temperature. 
a. Bulk modulus $K_T$ compared with previous work \cite{Sutton1953,Tallon1979}.
b. Relaxed volume $V_0$ of the simulation cell (at finite temperature and zero pressure). 
c. Volume thermal expansion coefficient $\alpha_V$ compared with work from the literature \cite{Touloukian1975}.
d. Deduced lattice parameter $a_{Al}$ as a function of the temperature.}
\label{Figalat}
\end{figure}

The lattice parameter $a_{al}(T)$ is determined at finite temperature by calculating the free energy $F(V,T)$ of thz perfect crystal having a hydrostatic deformation, which is defined as:
\begin{equation}
	F(V,T) = E(V)+F_{vibr}(V,T),
\label{eqAF}
\end{equation}

with $E(V)$, the static energy and $F_{vibr}(V,T)$ the contribution from the vibrations of the crystalline lattice calculated with the quasi-harmonic approximation (QHA) using phonon frequencies $\omega_j$ \cite{Metsue2011}:
\begin{multline}
\label{eqAFvibr}
	F_{vibr}(V,T) = \frac{1}{2}\sum_{q,j}h\omega_j(\vec{q},V) \\
	+k_BT\sum_{\vec{q},j}\ln{2\sinh{\left[\frac{h\omega_j(\vec{q},V)}{2k_BT}\right]}},
\end{multline}

with the first and second terms, the zero-point energy and thermal contributions, respectively.
The vibration frequencies of phonons, $\omega_j$ are the eigenvalues of the dynamic matrix for a wave vector $\vec{q}$.
The pressure $P$ is calculated for several volumes $V$, and the pressure variation depending on the volume is fitted with the Vinet equation of states \cite{Vinet1987,Hachet2018}: 
\begin{multline}
\label{eqAEoS}
P(V) = 3K_0 \frac{1-(V/V_0)^{1/3}}{(V/V_0)^{2/3}}\times \\
\exp\left\lbrace\frac{3}{2}(K'_0-1)\lbrack1-(V/V_0)^{1/3}\rbrack\right\rbrace,
\end{multline}

with $K_0$ the bulk modulus equal and $K'=\lbrack\partial K_0(T)/\partial P\rbrack_T$.
The lattice parameter $a_{Al}$ is obtained from $V_0$ the volume of the cell at 0\,GPa and at finite temperature.
Fig. \ref{Figalat}.a presents the isothermal bulk modulus $K_T$, compared with results from the literature \cite{Sutton1953,Tallon1979}. 
These calculations slightly underestimate the constant, which is also observed in nickel \cite{Hachet2018}. 
However, the difference, which could come from the exchange-correlation potential \cite{Lejaeghere2014} remained below 10\,\%. 
From the relaxed $V_0$ (fig. \ref{Figalat}.b), the volume thermal expansion coefficient $\alpha_V$ (fig. \ref{Figalat}.c) is determined and is close to the values available in the literature \cite{Touloukian1975}. 
Then, the lattice parameter $a_{Al}$ is deduced and presented in fig. \ref{Figalat}.d. 

\bibliographystyle{elsarticle-num-names}
\biboptions{sort&compress}
\bibliography{BibActaMat}

\end{document}